\documentclass[12pt]{article}

\usepackage{amsmath,amssymb,amsthm,graphicx}

\newcommand{\ed}{\end{document}}
\newcommand{\be}{\begin{equation}}
\newcommand{\ee}{\end{equation}}

\begin{document}
\begin{center}
\large{\textbf{\textbf{Black Hole Entropy: From Shannon to Bekenstein}}}\\
\end{center}
\begin{center}
Subir Ghosh\\
Physics and Applied Mathematics Unit, Indian Statistical
Institute\\
203 B. T. Road, Kolkata 700108, India \\
\end{center}\vspace{1cm}

\begin{center}
{\textbf{Abstract}}
\end{center}
In this note we have  applied directly the Shannon formula for information
theory entropy to derive
the Black Hole (Bekenstein-Hawking) entropy. Our analysis is  semi-classical in
nature since we use
the (recently proposed [8]) quantum mechanical near horizon mode functions to
compute the tunneling probability that goes in to
the Shannon formula, following the general idea of [5]. Our framework conforms
to the information theoretic origin of Black Hole entropy,
as originally proposed by Bekenstein.
\vspace{1cm}

Black Holes (BH) are, quite paradoxically, the simplest objects to
describe (thermodynamically) in terms of Bekenstein-Hawking
entropy, Hawking temperature etc. \cite{h,b} due to the ``no hair'' theorems
leading to their universal characters, but  at the same time, they
are probably one the most complex systems if one tries to
understand (in a statistical mechanical way)  the microscopic
degrees of freedom (or equivalently quantum states) that are
responsible for the huge BH entropy. In fact, the information
theoretic origin of BH entropy was strongly emphasized by
Bekenstein in the seminal work \cite{b}. Quantitative estimation
of information, or its loss in terms of information theoretic
``entropy'', was first defined in the pioneering work of Shannon
\cite{sh}. Bekenstein, in \cite{b}  refers to Shannon's expression
for entropy in his heuristic derivation of the BH entropy, but
explicitly uses it only to fix the normalization of the BH entropy
unit, a $bit$. One $bit$ of information is associated with the
knowledge of existence of just  one particle and $1~bit$ corresponds to $ln~2$ of
entropy. This
easily follows from the Shannon formula \cite{sh} for entropy $H$,
\be H=-\sum _ k p_k~ ln~p_k, \label{s} \ee where in a very
general context, $p_k$ denotes the probability  of the system
being in $k$-th state. For the particle entropy, it is numerically
equal to maximum entropy $ln~2$ if the chances of the particle
existing or not are both equal to $1/2$. Hence it was not clear if
the Shannon formula was capable of giving the BH entropy. Also it
is important to note, (as was stressed by Bekenstein \cite{b}), a
quantum treatment in evaluating the BH entropy seemed inevitable
but unfortunately the requisite results were not available at that
time.

The connection between Shannon entropy and thermodynamic or
statistical mechanics entropy was established by the works of
Jaynes \cite{j} and Brillouin \cite{br}. More recently,
Bialynicki-Birula and Mycielski \cite{bb} have exploited these
ideas to introduce new forms of ``entropic uncertainty relations''
applicable to physics problems, that use Shannon entropy forms. In
the present brief note we will follow \cite{bbr} where the authors
advocate the use of conventional quantum mechanical probability
expressions to compute the Shannon entropy. Our aim is to compute
the BH entropy using Shannon formula in a semi-classical
framework. This will vindicate the early usage of information
theory ideas in deriving BH entropy by Bekenstein and equally
important, we will give a quantum mechanical, (at least
semi-classical), treatment which although mentioned, was absent in
\cite{b}. This missing link has been given, very recently, by
Banerjee and Majhi in \cite{bm} where they re-discover the explicit
forms of quantum mechanical modes of the states inside and outside
the BH horizon (see also \cite{pc}). This directly yields the probabilities of
individual ingoing modes being trapped inside the BH horizon,
(which is indeed unity), or tunneling out of the BH horizon and
escaping to infinity to be perceived as Hawking radiation. The
algebraic structure of these probability expressions are
tailor-made for our purpose as they provide a back of the envelope
derivation of BH entropy via Shannon formula. These quantum modes have been
used
recently \cite{doug} to show the possibility of  non-divergent behavior for BH
(Hawking) temperature
as BH mass goes to zero and to indicate  a possible solution to the information
puzzle.

We follow the notations and conventions of \cite{bm}. In a very important paper
by Robinson and Wilczek \cite{wil} it was observed that effective field theories
become two dimensional and chiral near the event horizon of a  black hole. Explicitly, upon transforming to the  ``tortoise''
coordinate  and performing the partial wave decomposition, it can be shown that the effective radial potentials
for partial wave modes of the scalar field vanish expo-
nentially fast near the horizon. Thus physics near the
horizon can be described using an infinite collection of
(1 + 1)-dimensional fields. This approach has been used in various works 
\cite{others}.
This simplification allows us consider, in a BH background, the near horizon
quantum dynamics of the degrees of
freedom to be essentially two dimensional and the reduced metric in
$r-t$ sector, for a spherically symmetric and static spactime metric  is, \be
ds^2=F(r)dt^2 -\frac{dr^2}{F(r)}-r^2d\Omega ^2.
\label{bh} \ee For Hawking radiation the radial motion is relevant.
The massless Klein-Gordon equation in the above reduced metric
is, \be
g^{\mu\nu}\nabla _\mu\nabla _\nu \phi =0.
\label{kg} \ee One uses the WKB ansatz to solve the equation in a
semi-classical Hamilton-Jacobi scheme. It is convenient to use
tortoise coordinates and one has to keep in mind the change in the
coordinates as the outgoing particle crosses the horizon (for details see
\cite{bm}).
Explicit expressions of the modes are
\begin{eqnarray}
\label{modes}
\phi_{\text{in}}^{\text{(R)}}&=& \frac{1}{\sqrt{4\pi \omega
}}e^{-\frac{i}{\hbar}\omega
\,u_{\text{in}}}=\frac{1}{\sqrt{4\pi \omega }} e^{-\frac{\pi
\omega}{\hbar\kappa}}~e^{-\frac{i}{\hbar}\omega
\,u_{\text{out}}}= e^{-\frac{\pi
\omega}{\hbar\kappa}}~\phi_{\text{out}}^{\text{(R)}}\nonumber\\
\phi_{\text{in}}^{\text{(L)}}&=& \frac{1}{\sqrt{4\pi \omega }}
e^{-\frac{i}{\hbar}\omega
\,v_{\text{in}}}= \frac{1}{\sqrt{4\pi \omega }}e^{-\frac{i}{\hbar}\omega
\,v_{\text{out}}}=\phi_{\text{out}}^{\text{(L)}}.
\end{eqnarray}
In the above $\omega$ is the energy of the particle as measured by an asymptotic
observer and $\kappa$ is the surface gravity.  The null tortoise coordinates are
defined as  $u=t-r_\star$, $v=t+r_\star$
 $\left(r_\star =\int\frac{dr}{F(r)}\right)$
corresponding to the $2-d$ metric $ds^2=-F(r)dt^2+\frac{dr^2}{F(r)}$. Left (L)
moving modes are ingoing and Right (R) moving modes are  outgoing, with respect
to the horizon. Notice that the modes are expressed in terms of an outside
observer. The two-dimensional coordinate system in the null tortoise coordinates
is locally
conformally flat and hence the modes are  normalized in the conventional flat
spacetime way with an integral over a constant-time hypersurface ${\Sigma _t}$\\
\be
\mid \phi ^*\phi \mid =-i\int _{\Sigma _t} (\phi \partial _t\phi ^*-\phi^*
\partial _t\phi )dx~;~~
\int e^{i\vec k.\vec x}dx=2\pi\delta (\vec k).
\ee

Now, with the probabilities normalized as above, probability of  the left moving
$L$ modes to travel
towards the centre of
the black
hole should be unity. Clearly this is the case since \eqref{modes},
\be P^{\text{(L)}}=\left\vert \phi_{\text{in}}^{\text{(L)}}
\right\vert^2=\left\vert \phi_{\text{out}}^{\text{(L)}} \right\vert^2=1.
\label{mode} \ee
On the other hand, probability for the right moving $R$ modes to tunnel out
through the
  horizon is,
 \begin{eqnarray}
\label{outProbab}
P^{\text{(R)}}=\left\vert \phi_{\text{in}}^{\text{(R)}} \right\vert^2=\left\vert
e^{-\frac{\pi \omega}{\hbar\kappa}}\,\phi^{\text{(R)}}_{\text{out}}\right\vert^2
=e^{-\frac{2\pi \omega}{\hbar\kappa}},
\end{eqnarray}
as measured by an external observer. Utilizing the principle of detailed
balance $P^{\text{(R)}}=e^{-\frac{\omega}{T}}~P^{\text{(L)}}$ one recovers the
Hawking temperature $T=\frac{\hbar\kappa}{2\pi}$ \cite{bm} (see also discussions
in \cite{doug1}).

 Armed with these inputs we are now ready to exploit the idea of
Bialynicki-Birula and  Rudnicki \cite{bbr} to recover the
BH entropy from Shannon formula . For the present case  (\ref{s}) reduces to,
 \be
H=-\int_0^\infty \frac{d\omega }{cE_P}~P(\omega )~ln P(\omega ) =-\int_0^\infty
\frac{d\omega }{cE_P}~exp(-\alpha \omega )(-\alpha \omega ),
\label{sen}
\ee
where we have scaled the energy-like variable $\omega $ by 
$cE_P=c\sqrt{(\hbar c^5)/G}$, with $E_P$ being the Planck energy, to make the
Shannon entropy
dimensionless and $\alpha =(2\pi )/(\hbar\kappa)=(8\pi GM)/(\hbar c^4)$ from
(\ref{outProbab})
\cite{bm}. Here $\kappa =(c^4)/(4GM)$ is the surface gravity with $M$ being the
Black Hole mass. We have also reverted back to conventional $c.g.s.$ system of
units. A simple integration yields the result,
\be
H=\frac{1}{cE_P\alpha }=\frac{1}{8\pi (M/M_P)},
\label{h1}
\ee
with $M_P=\sqrt{(\hbar c)/G}$ denoting the Planck mass.
 Note that only the probability for
outgoing modes contribute since $P^{(L)}(\omega )=1$ for ingoing modes
which when substituted in Shanon formula yields zero.

Let us interpret this result in its proper perspective. We will attempt to
relate this result
to the  entropy problem as perceived by an observer
outside the BH horizon. Consider that a pair of left moving and  right moving
modes
is created just outside the BH horizon. We would like to reduce the problem to
that of a yes-no
type of two state problem, that is simply whether a particle exists or not.
According to the observer the right moving mode
(that tunnels out) exists and hence contributes zero entropy. For the left
moving mode that goes
inside the horizon, we fall back on Bekenstein argument \cite{b} and claim that
we do not really know anything about
the particle's existence or equivalently that it has equal ($\frac{1}{2}$)
probability of existing or not
existing. This means it is in a random state and can contribute the maximum
entropy $1~bit$. This is also justified since it is a well known result that the
mean
entropy carried away by a  mode having thermal distribution  is of the order of
one $bit$.  We put this as the upper bound of the amount of information that
is lost by the BH ($\frac{1}{8\pi (M/M_P)}$) from within the
horizon as a particle tunnels out.

As we have mentioned, our aim is to relate the two expressions of entropy, 
(\ref{h1}) obtained from the black hole point of view and depending on black
hole parameter $M$ to the entropy considered just above, from a very general
degree of freedom perspective. This in turn means that $\frac{1}{(M/M_P)}$
amount of BH mass (in Planck units) corresponds to $(8\pi)ln 2$
amount of (Shanon) entropy. Then an
$M/M_P$-mass BH will correspond to $ (8\pi)ln 2 (M/M_P)^2$ amount
of (Shannon) entropy
or more precisely
\be
S_{BH}\leq (ln 2)8\pi (M/M_P)^2 k_B.
\label{enbh}
\ee
In the above, we have denoted $S_{BH}$ to be the thermodynamic entropy which is
obtained by multiplying
the Shannon enropy by $k_B$, the Boltzmann's constant.  The above saturation
result,
\be
S_{BH}=(ln 2)8\pi
(M/M_P)^2 k_B,
\label{enbh1}
\ee
 in terms of the BH area $ A=4\pi
R_S^2=4\pi
(\frac{2MG}{c^2})^2$ (for a Schwarzschild BH), reads
\be
S_{BH}=\frac{8\pi(ln 2)}{M_P^2}\frac{c^4A}{16\pi G^2}k_B
= (\frac{ln2}{2}k_Bc^3\hbar ^{-1}G^{-1}A).
\label{bek}\ee
Notice that after
resurrecting the parameters in terms of fundamental constants
our result looks
tantalizingly close to the famous BH entropy expression of Bekenstein \cite{b}
$$S_{BH}=(ln2/4\pi)k_Bc^3\hbar ^{-1}G^{-1}A,$$ both expressionwise as well as
numerically.
Indeed it will be worthwhile to
see whether our argument and the final result (\ref{bek}) can refined further.

To conclude, we have attempted to apply directly the Shannon information
theoretic formula to derive the Black Hole entropy. This is along
the idea of Bekenstein  and we quote from \cite{b}: ``At the outset it should be
clear that the black hole entropy we are speaking of is {\it{not}} the thermal
entropy inside the black hole.'' We have shown that it can be directly obtained
from the Shannon formula of information entropy.
Furthermore Bekenstein also emphasizes that a quantum analysis is necessary to
account for the appearance
of $\hbar$ in the entropy expression. In the present work we have used
probability expressions obtained
from wave functions derived very recently \cite{bm} in a semiclassical
framework.
The application of quantum results in Shannon's
formula was outlined in \cite{bbr}. We have taken Schwarzschild BH solution as
an example but our analysis can be easily
extended to generic forms of Black Holes. One only needs to substitute the
appropriate expression for the surface gravity   in the mode functions
$\phi^{(R)}_{in}$ and $\phi^{(L)}_{in}$.

There are three  interesting directions along which the present analysis can be
extended. One is the application of  ``Entropic''
Uncertainty Relations, developed by Bialynicki-Birula and co-workers,
\cite{bb,bbr}, in the present context. In a similar vein, extensions
of laws to include quantum corrections due to conventional Heisenberg
Uncertainty Relation and Generalized Uncertainty Relation have appeared in
\cite{v} in the context of
``entropic'' origin of gravity, recently proposed by Verlinde \cite{ve}. This
actually brings us to the second direction of future work where
it would be very interesting to derive rigorously Verlinde's conjecture
\cite{ve} that relates the entropy change in a holographic screen
due to a particle's motion to its distance from the screen. Indeed it seems
natural that Shannon's entropy formula should come in to play since
it was stressed by Verlinde \cite{ve} that the origin of gravity is entirely
information theoretic. Since Verlinde's entropic analysis of Newton's
law of gravitation resides in flat spacetime, the global embedding procedure,
recently developed in \cite{bm1}, that studies the Hawking effect
 (curved spacetime) and Unruh effect \cite{u} (Minkowski spacetime), can play an
important role. The third is  to apply the line of argument presented here in
obtaining entropies for more general type of black holes, such as Kerr black
hole in particular.
\vskip .3cm
{\it{Acknowledgements}}: It is a pleasure to thank Professor Rabin Banerjee and
Professor Guruprasad
Kar for
discussions. I am also grateful to Professor Jacob D. Bekenstein  and Professor
Samuel L. Braunstein for helpful correspondences.

\end{document}